\documentclass{article}

\usepackage{epsfig} 
\newlength{\abstractwidth} 
\renewcommand{\thefootnote}{\fnsymbol{footnote}} 
\renewcommand{\thanks}[1]{\footnote{#1}} 
\newcommand{\starttext}{ 
\setcounter{footnote}{0} 
\renewcommand{\thefootnote}{\arabic{footnote}}} 
\renewcommand{\theequation}{\thesection.\arabic{equation}} 
\newcommand{\bea}{\begin{eqnarray}} 
\newcommand{\eea}{\end{eqnarray}} 
\newcommand{\beq}[1]{\begin{equation} \label{#1}} 
\newcommand{\be}{\begin{equation}} 
\newcommand{\ee}{\end{equation}} 
\newcommand{\half}{${1\over 2} \:$}

\def\12{{1 \over 2}}

\newcommand{\tab}{\hspace{5mm}}

\begin{document} 
\renewcommand{\theequation}{\thesection.\arabic{equation}} 
\begin{titlepage} 
\bigskip
\centerline{\Large \bf {Linear Spinor Field Equations for Arbitrary Spins}}
\bigskip
\begin{center} 
{\large James Lindesay\footnote{
Permanent address, Department of Physics, Howard University, Washington, DC 20059}
} \\
\end{center}
\bigskip\bigskip 
\begin{abstract} 

When utilizing a cluster decomposible relativistic scattering formalism, it
is most convenient that the covariant field equations take on a linear form
with respect to the energy and momentum dispersion on the fields in the
manner given by the Dirac form for spin \half  systems.  A general spinor
formulation is given for arbitrary spins by minimally extending the Lorentz
algebra to include operators whose matrix representation give general
Dirac matrices.  The forms of these matrices are explicitly demonstrated for spin
\half and spin 1 fields.

\end{abstract} 
\end{titlepage} 


\starttext \baselineskip=17.63pt \setcounter{footnote}{0} 
\setcounter{equation}{0} 
\section{ Introduction} 
\tab 

The Dirac equation utilizes a matrix algebra to construct a linear operator
relationship between the energy and the momentum.  When constructing
a scattering formalism for relativistic quantum systems, the off-diagonal
(off-shell) nature of the intermediate states considerably complicates the
implementation of the separation of the purely kinematic variables of the
non-interacting components of a multi-particle system from the dynamical
variables of the interacting components, due to the generally
non-linear dispersion relationships between
energies and momenta.  One is able to demonstrate the general cluster
decomposibility of a multiparticle system if the off-shell behavior is parametric,
and the resolvants satisfy linear dispersions\cite{LMNP}\cite{AKLN}\cite{Compton},
as is the case using the resolvants resulting from the Dirac equation.  This
makes the development of linear dispersion equations for systems of
arbitrary spin crucial for the implementation of physical cluster decomposibility
in general, non-perturbative, formal scattering theory.

It is therefore advantageous to extend the Lorentz group to include operators
whose matrix elements reduce to the Dirac matrices for spin $1 \over 2$
systems, but generally require that the form $\hat{\Gamma}^\mu \: \hat{P}_\mu$
be a scalar operation.  The form $\hat{\Gamma}^\mu$ will be seen to be
associated with the particle type current,
where $\hat{P}_\mu$ is associated with the mass-energy current (space-time
translations).  For the present purposes, we will assume that the group
is internal, commuting with global space-time translations.  In a subsequent
paper we will present the extended Poincare group behavior for systems
that transform under this group structure.  The finite dimensional representations
of the extended Poincare group will there be
constructed in a manner analogous to the use of the Little group on
the standard state vectors in the construction of finite dimensional
representations for spin and helicity in the Poincare algebra.
\beq{transformation}
\hat{U}(\underline{b}) \, \left | \psi_\lambda \: \vec{a} \right \rangle \: = \:
\sum_{\lambda'}^{} \, \left | \psi_\lambda' \: \vec{z}(\underline{b}; \vec{a}) \right \rangle
\, {\mathit{D}^{\lambda'}}_\lambda (\underline{b}; \vec{a} )
\ee

A single-particle wave equation can be developed for configuration space
eigenstates of the operator $\hat{\Gamma}^\mu \: \hat{P}_\mu$,
\be
\mathbf{\Gamma}^\mu \: i \partial_\mu \psi (x) \: = \: \lambda \: \psi (x)
\ee
which implies that
\be
\partial_\mu \left [ \overline{\psi (x)} \: \mathbf{\Gamma}^\mu \: \psi (x) \right ] \: = \: 0.
\label{current}
\ee
The conserved current defined in Equation \ref{current} need not be the
probability current, since the spinor metric need not be related to the
$\Gamma$ matrices in general.  However, scattering equations can be
developed to express the evolution of whatever physical parameter
is represented by this operator.

In what follows, the group structure of an extended Lorentz algebra
will be developed.  Finite dimensional representations of this group
will be expressed in terms of spinors, and matrix representations
will be developed for two systems of interest.

\setcounter{equation}{0} 
\section{Extended Lorentz Group} 

\tab 
The finite dimensional representations of this extended group will be
constructed by developing a spinor representation of the algebra.
The group elements $\underline{b}$ will for the present include
3 parameters representing angles, 3 boost parameters, and 4
group parameters $\vec{h}$ associated with the operators $\Gamma$.  
Later, it can be expanded to include at least the fourteen parameters
including the space-time translations $\vec{b}$. 
For the present, the extended Lorentz group sub-algebra will be developed.

\subsection{Extended Lorentz Group Commutation Relations} 
The extended group commutation relations will be chosen to be consistent
with the Dirac matrices as follows:
\bea
\left [ J_j \, , \, J_k \right] \: = \: i \, \epsilon_{j k m} \, J_m \\
\left [ J_j \, , \, K_k \right] \: = \: i \, \epsilon_{j k m} \, K_m \\
\left [ K_j \, , \, K_k \right] \: = \: -i \, \epsilon_{j k m} \, J_m \\
\left [ \Gamma^0 \, , \, \Gamma^k \right] \: = \: i \, K_k \\
\left [ \Gamma^0 \, , \, J_k \right] \: = \: 0 \\
\left [ \Gamma^0 \, , \, K_k \right] \: = \: -i \,  \Gamma^k \\
\left [ \Gamma^j \, , \, \Gamma^k \right] \: = \: -i \, \epsilon_{j k m} \, J_m \\
\left [ \Gamma^j \, , \, J_k \right] \: = \: i \, \epsilon_{j k m} \, \Gamma^m \\
\left [ \Gamma^j \, , \, K_k \right] \: = \: -i \, \delta_{j k} \, \Gamma^0 
\eea
For convenience, define $\Delta_k ^{(\pm)}$ as follows:
\be
\Delta_k ^{(\pm)} \: \equiv \: \Gamma^k\, (\pm) \, i  K_k  .
\ee
Then the algebra can be expressed in terms of raising and lowering operators
\bea
\left [ \Gamma^0 \, , \, J_k  \right] \: = \: 0 
\label{firstcommutator}\\
\left [ \Gamma^0 \, , \, \Delta_k ^{(\pm)} \right] \: = \: (\pm) \, \Delta_k ^{(\pm)} \\
\left [ J_z \, , \, J_\pm \right] \: = \: \pm \, J_\pm \\
\left [ J_+ \, , \, J_- \right] \: = \: 2 \, J_z \\
\left [ J_z \, , \, \Delta_z ^{(\pm)} \right] \: = \: 0 \\ 
\left [ J_z \, , \, \Delta_\pm ^{(\pm)} \right] \: = \: \pm \, \Delta_\pm ^{(\pm)} \\
\left [ J_\pm \, , \, \Delta_\pm ^{(\pm)} \right] \: = \: 0 \\
\left [ J_\pm \, , \, \Delta_\mp ^{(\pm)} \right] \: = \: \pm 2 \, \Delta_z ^{(\pm)} \\
\left [ J_\pm \, , \, \Delta_z ^{(\pm)} \right] \: = \: \mp \, \Delta_\pm ^{(\pm)} \\
\left [ \Delta_z ^{(+)} \, , \, \Delta_z ^{(-)} \right] \: = \: -2 \, \Gamma^0 \\
\left [ \Delta_z ^{(\pm)} \, , \, \Delta_\pm ^{(\pm)} \right] \: = \: 0 \\
\left [ \Delta_z ^{(+)} \, , \, \Delta_\pm ^{(-)} \right] \: = \: \mp 2 \, J_\pm \\
\left [ \Delta_z ^{(-)} \, , \, \Delta_\pm ^{(+)} \right] \: = \: \mp 2 \, J_\pm \\
\left [ \Delta_+ ^{(\pm)} \, , \, \Delta_- ^{(\pm)} \right] \: = \: 0 \\
\left [ \Delta_+ ^{(+)} \, , \, \Delta_+ ^{(-)} \right] \: = \: 0 \\
\left [ \Delta_+ ^{(\pm)} \, , \, \Delta_- ^{(\mp)} \right] \: = \: -4 \left[ J_z (\pm) \Gamma^0 \right]
\label{lastcommutator}
\eea
which will be a convenient form from which to construct the spinor representation.

\subsection{Group metric}

In general, a group metric can be developed from the adjoint representation
in terms of the structure constants.  For the algebra represented by
\be
\left [ \hat{G}_r \, , \, \hat{G}_s \right ] \: = \: -i \, \left ( g_s \right ) _r ^m \, \hat{G}_m
\ee
the group metric can be defined by
\be
\eta_{a b} \: \equiv \: \left ( g_a \right )_r ^s \, \left ( g_b \right )_s ^r .
\ee
The non-vanishing components of the extended Lorentz group metric are
given by
\bea
\eta^{(EL)} _{J_m \, J_n} \: = \: -6 \, \delta_{mn} \quad , \quad &
\eta^{(EL)} _{K_s \, K_n} \: = \: +6 \, \delta_{mn} \quad , \quad &
\eta^{(EL)} _{\Gamma^\mu \, \Gamma^\nu} \: = \: +6 \, \eta_{\mu \nu}
\eea
It is interesting to note that the group structure of the extended Lorentz
group generates the Minkowski metric.  Neither the group structure of
the usual Lorentz group nor that of the Poincare group can generate
the Minkowski metric due to the abelian nature of the generators for
infinitesimal space-time translations.

\setcounter{equation}{0} 
\section{Spinor Equations} 

A Casimir operator can be constructed for the extended Lorentz group, 
given by
\be
C \: = \: \underline{J} \cdot \underline{J} \,-\, \underline{K} \cdot \underline{K}
\,+\, \Gamma^0 \, \Gamma^0 \,-\, \underline{\Gamma} \cdot \underline{\Gamma} .
\ee
This operator can directly be verified to commute with all generators of the group. 
We will construct eigenstates of this Casimir operator, along with the commuting
operators $J_z$ and $\Gamma^0$.  To develop a basis of states,
it is convenient to construct an operator which raises and lowers eigenvalues
of the operator $\Gamma^0$ analogous to the angular momentum raising and
lowering operators.  This operator is given by
\be
\Delta_J ^{(\pm)} \: \equiv \: \underline{J} \cdot \underline{\Delta}^{(\pm)}
\ee
The relavant commutation relations needed to construct the basis are given by
\bea
\left [ J_z \, , \, J_\pm \right] \: = \: \pm \, J_\pm \\
\left [ \Gamma^0 \, , \, \Delta_J ^{(\pm)} \right] \: = \: (\pm) \, \Delta_J ^{(\pm)} \\
\left [ J_z \, , \, \Delta_J ^{(\pm)} \right] \: = \: 0 \\
\left [ \Gamma^0 \, , \, J_\pm \right] \: = \: 0 \\
\left [ J_z \, , \, \Gamma^0 \right] \: = \: 0 \\
\left [ J^2 \, , \, \underline{J} \right] \: = \: 0 \\
\left [ J^2 \, , \, \Gamma^0 \right] \: = \: 0 \\
\left [ J^2 \, , \, \Delta_J ^{(\pm)} \right] \: = \: 0
\eea

\subsection{Conjugate Spinor Forms}
We can construct conjugate spinor forms for the operators that satisfy
Equations \ref{firstcommutator}-\ref{lastcommutator}.
\bea
J_z \: = \: {1 \over 2} \left[ \chi_{(+)} ^+ \partial_+ ^{(+)} \,+\,
\chi_+ ^{(-)} \partial_+ ^{(-)} \,-\,
\chi_- ^{(+)} \partial_- ^{(+)} \,-\,
\chi_- ^{(-)} \partial_- ^{(-)} \right]
\label{firstspinorform} \\
J_\pm \: = \: \chi_\pm ^{(+)} \partial_\mp ^{(+)} \,+\,
\chi_\pm ^{(-)} \partial_\mp ^{(-)} \\
\Gamma^0 \: = \: {1 \over 2} \left[ \chi_{(+)} ^+ \partial_+ ^{(+)} \,-\,
\chi_+ ^{(-)} \partial_+ ^{(-)} \,+\,
\chi_- ^{(+)} \partial_- ^{(+)} \,-\,
\chi_- ^{(-)} \partial_- ^{(-)} \right] \\
\Delta_z ^{(\pm)} \: = \: (\pm) \left[ \chi_+ ^{(\pm)} \partial_+ ^{(\mp)} \,-\,
\chi_- ^{(\pm)} \partial_- ^{(\mp)} \right] \\
\Delta_\pm ^{(\pm)} \: = \: (\pm) \, 2 \, \chi_\pm ^{(\pm)} \partial_\mp ^{(\mp)}
\label{lastspinorform}
\eea
This representation provides a convenient mechanism to construct
spinor and matrix representations.

\subsection{Symmetry Behavior of Spinor Forms}

Examine the behavior of the operators under the transformation given by
\be
\chi_\pm ^{(\pm)} \: \leftrightarrow \: \bar{\chi}_\pm ^{(\mp)} .
\ee
The angular momentum and $\Gamma^0$ operators can be seen to transform as
\bea
\underline{J} \: \leftrightarrow \: \bar{\underline{J}} \\
\Gamma^0 \: \leftrightarrow \: -\bar{\Gamma}^0
\eea
The commutation relations are preserved for the various generators
in the bar representation.  For the Dirac case, this will be seen to
represent a "particle-antiparticle" symmetry of the system, and it will
represent a general symmetry under negation of the eigenvalues of
the operator $\Gamma^0$.

\subsection{General Construction of States}
We can next construct general spinor states by operation of the raising
operators $\Delta_J ^{(+)}$ and $J_+$ on the minimal state
$\psi_{\gamma_{min} , -J} ^{\Lambda , J}$.  This gives a general form for
$\psi_{\gamma , M} ^{\Lambda , J}$ given by
\be
\begin{array}{r}
\psi_{\gamma , M} ^{\Lambda , J} \: = \: A^{\Lambda J}
\sqrt{{(J-M)! \over (J+M)! \, (2J)!}} \,[x-y]^{\Lambda - J} \, 
\chi_+ ^{(+) \, M+\gamma} \, \chi_+ ^{(-) \, M-\gamma} \,\times \\
\\
\left .
\left[ {\partial \over \partial x} \,+\, {\partial \over \partial y}   \right] ^{J+M}
x^{J-\gamma} \, y^{J+\gamma} \right | _{
\begin{array}{l}
x=\chi_+ ^{(+)} \, \chi_- ^{(-)} \\ y=\chi_- ^{(+)} \, \chi_+ ^{(-)} 
\end{array} }
\end{array}
\ee
The action of the spinor forms of the operators given in Equations \ref{firstspinorform}-
\ref{lastspinorform} results in the following set of equations:
\bea
\hat{J}^2 \, \psi_{\gamma , M} ^{\Lambda , J} \: = \: 
J(J+1) \,\psi_{\gamma , M} ^{\Lambda , J} \quad , \quad &
\hat{C} \, \psi_{\gamma , M} ^{\Lambda , J} \: = \: 
2 \Lambda (\Lambda+2) \,\psi_{\gamma , M} ^{\Lambda , J} \\
\hat{J}_z \, \psi_{\gamma , M} ^{\Lambda , J} \: = \: 
M \,\psi_{\gamma , M} ^{\Lambda , J} \quad , \quad &
\hat{\Gamma}^0 \, \psi_{\gamma , M} ^{\Lambda , J} \: = \: 
\gamma \,\psi_{\gamma , M} ^{\Lambda , J} \\
\hat{J}_\pm \, \psi_{\gamma , M} ^{\Lambda , J} \: = \: 
\sqrt{(J \pm M + 1)(J \mp M)} \,\psi_{\gamma , M \pm 1 } ^{\Lambda , J} \quad , \quad &
\hat{\Delta}^{(\pm)} \, \psi_{\gamma , M} ^{\Lambda , J} \: = \: 
(\pm)\,(\Lambda+1)\,[J (\mp) \gamma] \,\psi_{\gamma \pm 1 , M} ^{\Lambda , J}
\label{op-spinor}
\eea
Equation \ref{op-spinor} clearly allows the construction of finite dimensional
representation if $J$ is included in the eigenvalue spectrum of $\Gamma^0$.

\subsection{Number of States}

The order of the spinor polynomial of the finite dimensional state with $\Lambda=J_{max}$
can be determined by examining the minimal state from which other
states can be constructed using the raising operators and orthonormality:
\be
\psi_{-\Lambda , -\Lambda} ^{\Lambda , \Lambda} \: = \: A^{\Lambda J} \chi_- ^{(-) 2 \Lambda} .
\ee
The general state involves spinor products of the order
\be
\chi_+ ^{(+) a} \chi_- ^{(+) b} \chi_+ ^{(-) c} \chi_- ^{(-) d}.
\ee
A complete basis of states requires then that $a+b+c+d=2 \Lambda$.  By direct counting
this yield the number of states for a complete basis:
\be
N_\Lambda \: = \: {1 \over 3} (\Lambda + 1) (2 \Lambda + 1) (2 \Lambda + 3).
\ee
For instance, $N_0 = 1, N_{1 \over 2} = 4, N_1 = 10, N_{3 \over 2} = 20$, and so on.

A single J basis with $(2 J + 1)^2$ states does not cover this space of spinors.  However,
one can directly verify that
\be
N_{J_{max}} \: = \: \sum_{J=J_{min}}^{J_{max}} (2 J + 1)^2 ,
\ee
where $J_{min}$ is zero for integral systems and $1 \over 2$ for half integral systems.
Thus we see that $\Lambda$ represents the maximal angular momentum state of the system:
\be
J \: \le \: \Lambda = J_{max} .
\ee

\subsection{Spinor metrics}

Invariant amplitudes are defined using dual spinors so that
under transformations the inner product is a scalar
\be
\begin{array}{c}
<\bar{\psi} | \phi > \: = \: <\bar{\psi'} | \phi'> \\
\psi_a ^\dagger g_{ab} \phi_b \: = \: 
\left( D_{ca} \psi_a  \right) ^\dagger g_{cd} 
\left( D_{db} \psi_b  \right)
\end{array}
\ee
This means the metric should satisfy
\be
\mathbf{g} \: = \: \mathbf{D^\dagger g D}
\ee

The eigenvalues of the angular momentum operator and $\Gamma^0$ will
be given by real numbers 
\be
\underline{J} ^\dagger=\underline{J}  \quad , \quad 
\Gamma^{0 \, \dagger}=\Gamma^0 
\ee
Since the spinor metric is likewise hermitian, it satisfies
\be
\mathbf{g \, \Gamma^0} \, = \, \mathbf{\Gamma^0 \, g} \quad , \quad
\mathbf{g \, \underline{J} } \, = \, \mathbf{\underline{J} \, g} 
\ee

From equation \ref{op-spinor} it can be seen that the representation is finite
dimensional if $\Lambda=J$.  Within the $\Lambda=J$ subspace,
we can assume the metric is proportional
to the identity matrix for angular momentum.  We can then generally show that
\be
\begin{array}{l}
\mathbf{g}_{M' \gamma' ; M \gamma} ^{\Lambda: J';J} \:= \:
g_\gamma ^{\Lambda J} \, \delta_{J' J} \delta_{M' M} \, \delta_{\gamma' \gamma} \\
\mathbf{\Delta}_{M' \gamma' ; M \gamma} ^{(\pm) \: \Lambda: J';J} \:= \:
\Delta_{J'M'; JM} ^{(\pm) \: \Lambda , \gamma} \, \delta_{\gamma' , \gamma (\pm)1}
\end{array}
\ee

If the condition $\mathbf{g \, \Delta_J ^{(\pm)} } = - \mathbf{ \Delta_J ^{(\pm)} \, g }$ is satisfied,
then the metric can be chosen such that
\be
g_\gamma ^{\Lambda J} \: = \: (-)^{\Lambda-\gamma}
\ee
By expanding the anticommutator $\{ \mathbf{g},\mathbf{\Delta_J ^{\pm}} \}$,
this then implies that the vector components of the boost generator and $\Gamma$ matrices
must satisfy
\be
\mathbf{g \, \underline{\Gamma} } \, = \,- \mathbf{\underline{\Gamma} \, g} \quad , \quad
\mathbf{g \, \underline{K} } \, = \, -\mathbf{\underline{K} \, g}
\ee

\subsection{Construction of $\Lambda={1 \over 2}$ Systems}

The forms of the matrices corresponding to $\Lambda={1 \over 2}$ are
expected to have dimensionality $N_{1 \over 2}=4$, and
can be expressed in terms of the Pauli spin matrices as shown below:
\be
\begin{array}{l l}
\mathbf{\Gamma^0} \,=\, {1 \over 2} \left( \begin{array}{r r}
\mathbf{1} & \mathbf{0} \\ \mathbf{0} & -\mathbf{1} \end{array} \right)
\,=\, {1 \over 2} \mathbf{g} \quad \quad &
\mathbf{\underline{J} } \,=\, {1 \over 2} \left( \begin{array}{r r}
\mathbf{\underline{\sigma} } & \mathbf{0} \\ \mathbf{0} & \mathbf{\underline{\sigma} } 
\end{array} \right) \\ \\
\mathbf{\underline{\Gamma}} \,=\, {1 \over 2} \left( \begin{array}{r r}
\mathbf{0} & \mathbf{\underline{\sigma} } \\ 
-\mathbf{\underline{\sigma} } & \mathbf{0} \end{array} \right) &
\mathbf{\underline{K}} \,=\, -{i \over 2} \left( \begin{array}{r r}
\mathbf{0} & \mathbf{\underline{\sigma} } \\ 
\mathbf{\underline{\sigma} } & \mathbf{0} \end{array} \right) \\ \\
\mathbf{\Delta_k ^{(+)}} \,=\,  \left( \begin{array}{c c}
\mathbf{0} & \mathbf{\sigma}_k \\ \mathbf{0} & \mathbf{0} \end{array} \right) &
\mathbf{\Delta_k ^{(-)}} \,=\, \left( \begin{array}{c c}
\mathbf{0} & \mathbf{0} \\ -\mathbf{\sigma}_k & \mathbf{0} \end{array} \right)
\end{array}
\ee
The $\Gamma^\mu$ matrices can directly be seen to be proportional
to a representation of the Dirac matrices\cite{Dirac}\cite{BjDrell} . 

\subsection{Construction of $\Lambda=1$ Spinor States}

The forms of the matrices corresponding to $\Lambda=1$ are
expected to have dimensionality $N_1=10$.  We choose the normalization
of the spinors to satisfy
\be
\overline{\left ( \chi_a ^{(A)} \chi_b ^{(B)} \right )} \left ( \chi_{a'} ^{(A')} \chi_{b'} ^{(B')} \right )
\: = \: {1 \over 2} \left ( \delta^{AA'} \delta_{aa'} \delta^{BB'} \delta_{bb'}+
 \delta^{AB'} \delta_{ab'} \delta^{BA'} \delta_{ba'} \right )
\ee
which results in a state normalization of the form
\be
\overline{\psi_{\gamma , M} ^{\Lambda , J} } \: \psi_{\gamma' , M'} ^{\Lambda , J'} \: = \:
\delta^{JJ'} \delta_{\gamma \gamma'} \delta_{MM'} .
\ee
The spinor states satisfying this normalization are given by
\be
\begin{array}{l}
\psi_{0 , 0} ^{1 , 0} \: = \: \chi_+ ^{(+)} \chi_- ^{(-)} - \chi_- ^{(+)} \chi_+ ^{(-)}  \\
\\
\psi_{1, 1} ^{1 , 1} \: = \: \chi_+ ^{(+) \, 2}  \\
\psi_{1, 0} ^{1 , 1} \: = \: \sqrt{2} \: \chi_+ ^{(+)} \chi_- ^{(+)}    \\
\psi_{1, -1} ^{1 , 1} \: = \: \chi_- ^{(+) \, 2}  \\
\\
\psi_{0, 1} ^{1 , 1} \: = \: -\sqrt{2} \: \chi_+ ^{(+)} \chi_+ ^{(-)}    \\
\psi_{0, 0} ^{1 , 1} \: = \: -\chi_+ ^{(+)} \chi_- ^{(-)} - \chi_- ^{(+)} \chi_+ ^{(-)}   \\
\psi_{0, -1} ^{1 , 1} \: = \: -\sqrt{2} \: \chi_- ^{(+)} \chi_- ^{(-)}    \\
\\
\psi_{-1, 1} ^{1 , 1} \: = \: \chi_+ ^{(-) \, 2}  \\
\psi_{-1 , 0} ^{1 , 1} \: = \: \sqrt{2} \: \chi_+ ^{(-)} \chi_- ^{(-)}    \\
\psi_{-1 , -1} ^{1 , 1} \: = \: \chi_- ^{(-) \, 2}
\end{array}
\label{Spin1states}
\ee

\subsection{Matrix Representation of $\Lambda=1$ Systems}
We finally construct matrix elements of the operators given by
Equations \ref{firstspinorform}-\ref{lastspinorform} using the states given in
Equation \ref{Spin1states} to obtain the matrix representation for
$\Lambda=1$ systems:
\be
\begin{array}{c c}
\mathbf{\Gamma^0} \,=\, \left ( \begin{array}{r r r r}
0 & \underline{0} & \underline{0} & \underline{0} \\
\underline{0} & \mathbf{1} & \mathbf{0} & \mathbf{0} \\
\underline{0} & \mathbf{0} & \mathbf{0} & \mathbf{0} \\
\underline{0} & \mathbf{0} & \mathbf{0} & \mathbf{-1} 
\end{array} \right ) &
\mathbf{g} \,=\, \left ( \begin{array}{r r r r}
-1 & \underline{0} & \underline{0} & \underline{0} \\
\underline{0} & \mathbf{1} & \mathbf{0} & \mathbf{0} \\
\underline{0} & \mathbf{0} & \mathbf{-1} & \mathbf{0} \\
\underline{0} & \mathbf{0} & \mathbf{0} & \mathbf{1} 
\end{array} \right ) \\ \\
\mathbf{J_z} \,=\, \left ( \begin{array}{c c c c}
0 & \underline{0} & \underline{0} & \underline{0} \\
\underline{0} & \mathbf{J}_z & \mathbf{0} & \mathbf{0} \\
\underline{0} & \mathbf{0} & \mathbf{J}_z & \mathbf{0} \\
\underline{0} & \mathbf{0} & \mathbf{0} & \mathbf{J}_z  
\end{array} \right ) &
\mathbf{J_\pm} \,=\, \left ( \begin{array}{c c c c}
0 & \underline{0} & \underline{0} & \underline{0} \\
\underline{0} & \mathbf{J}_\pm & \mathbf{0} & \mathbf{0} \\
\underline{0} & \mathbf{0} & \mathbf{J}_\pm & \mathbf{0} \\
\underline{0} & \mathbf{0} & \mathbf{0} & \mathbf{J}_\pm 
\end{array} \right ) \\ \\
\mathbf{\Delta_z ^{(+)}} \,=\, \left ( \begin{array}{c c c c}
0 & \underline{0} & \underline{0} & 2\underline{v}_z ^T \\
-\underline{v}_z & \mathbf{0} & \mathbf{J}_z & \mathbf{0} \\
\underline{0} & \mathbf{0} & \mathbf{0} & 2\mathbf{J}_z \\
\underline{0} & \mathbf{0} & \mathbf{0} & \mathbf{0} 
\end{array} \right ) &
\mathbf{\Delta_z ^{(-)}} \,=\, \left ( \begin{array}{c c c c}
0 & 2\underline{v}_z ^T & \underline{0} & \underline{0} \\
\underline{0} & \mathbf{0} & \mathbf{0} & \mathbf{0} \\
\underline{0} & -2\mathbf{J}_z & \mathbf{0} & \mathbf{0} \\
-\underline{v}_z & \mathbf{0} & -\mathbf{J}_z & \mathbf{0} 
\end{array} \right ) \\ \\
\mathbf{\Delta_\pm ^{(+)}} \,=\, \left ( \begin{array}{c c c c}
0 & \underline{0} & \underline{0} & 2\underline{v}_\mp ^T \\
-\underline{v}_\pm & \mathbf{0} & \mathbf{J}_\pm & \mathbf{0} \\
\underline{0} & \mathbf{0} & \mathbf{0} & 2\mathbf{J}_\pm \\
\underline{0} & \mathbf{0} & \mathbf{0} & \mathbf{0} 
\end{array} \right ) &
\mathbf{\Delta_\pm ^{(-)}} \,=\, \left ( \begin{array}{c c c c}
0 & 2\underline{v}_\mp ^T & \underline{0} & \underline{0} \\
\underline{0} & \mathbf{0} & \mathbf{0} & \mathbf{0} \\
\underline{0} & -2\mathbf{J}_\pm & \mathbf{0} & \mathbf{0} \\
-\underline{v}_\pm & \mathbf{0} & -\mathbf{J}_\pm & \mathbf{0} 
\end{array} \right )
\end{array}
\ee 
where
\be
\begin{array}{c c c}
\underline{v}_z \,\equiv \, \left( \begin{array}{c}
0 \\ 1 \\ 0 \end{array} \right) &
\underline{v}_+ \,\equiv \, \left( \begin{array}{c}
-\sqrt{2} \\ 0 \\ 0 \end{array} \right) &
\underline{v}_- \,\equiv \, \left( \begin{array}{c}
0 \\ 0 \\ \sqrt{2} \end{array} \right) \\ \\
\mathbf{J}_z \,\equiv \, \left( \begin{array}{c c c}
1 & 0 & 0 \\ 0 & 0 & 0 \\ 0 & 0 & -1 \end{array} \right) &
\mathbf{J}_+ \,\equiv \, \left( \begin{array}{c c c}
0 & \sqrt{2} & 0 \\ 0 & 0 & \sqrt{2} \\ 0 & 0 & 0 \end{array} \right) &
\mathbf{J}_- \,\equiv \, \left( \begin{array}{c c c}
0 & 0 & 0 \\ \sqrt{2} & 0 & 0 \\ 0 & \sqrt{2} & 0 \end{array} \right) 
\end{array}
\ee
Since not all of the generators are Hermitian in this representation,
the representation is seen to be finite dimensional, but not unitary, 
(the same as for the $\Lambda={1 \over 2}$ system).

\section{Acknowledgements}

The author wishes to acknowledge the support of Elnora Herod and
Penelope Brown during the intermediate periods prior to and after
his Peace Corps service (1984-1988), during which time the bulk of this work was
accomplished.  In addition, the author wishes to recognize the
hospitality of the Department of Physics at the University of Dar
Es Salaam during the three years from 1985-1987 in which a substantial portion of
this work was done.

\end{document}